\newtheorem{theorem}{Theorem}

\documentclass{article}
\usepackage{amsfonts}

\usepackage{sw20aip}

\newtheorem{acknowledgement}[theorem]{Acknowledgement}

\input{tcilatex}
\begin{document}

\title{Chaos and Non-Archimedean metric in the Bernoulli Map}
\author{Jesus San-Martin$^1\thanks{%
corresponding author. e-mail: jsm@apphys.uned.es}$, Oscar Sotolongo-Costa$%
^{2,3}$ \\
1- Dpto. Matematica Aplicada, Esc. Universitaria de Estadistica,\\
Universidad Complutense.\\
Av. Pta. de Hierro S/N. 28040 Madrid.\\
2- LCTDI, UNED, a.p. 60141, 28080 Madrid, Spain.\\
3- Fac. de Fisica, Universidad de La Habana, Habana 10400, Cuba.}
\maketitle

\begin{abstract}
Ultrametric concepts are applied to the Bernoulli Map, showing the
adequateness of the non-Archimedean metrics to describe in a simple and
direct way the chaotic properties of this map. Lyapunov exponent and
Kolmogorov entropy appear to find a simpler explanation. A p-adic time
emerges as a natural consequence of the ultrametric properties of the map.

PACS: 05.45.+b, 02.90.+p

Keywords: Non-Archimedean, ultrametricity, chaos
\end{abstract}

\section{\protect\smallskip Introduction}

After the work by M\'{e}zard et al. \cite{mezard}, ultrametricity has
triggered the interest in a wide range of physical phenomena, due to its
applications in different topics: spin glasses, mean field theory,
turbulence, nuclear physics . Also optimization theory, evolution, taxonomy,
protein folding benefits from it ( for an excellent review see \cite
{rammal,brekke}). Wherever a hierarchical concept appears, non-Archimedean
analysis is an adequate tool to study the problem.

Ultrametricity is a promising tool in the theory of branching processes,
which also has revealed its possibilities in the study of self-organized
critical processes \cite{stanley,pelayo,prl}. It seems possible to find
simpler tools to describe the geometry of these processes. Here, we expose
the advantages of a hierarchical representation in the case of the Bernoulli
shift. This will permit, using simple geometric considerations, to determine
the magnitudes governing the system, and the advantages of a p-adic metric
will be stressed over the euclidean one. The ultrametric distance will be
shown to be consistent with the characteristic behavior of this chaotic
unidimensional map. In ultrametric spaces, concepts such as exponential
separation of neighboring trajectories, and characteristic parameters
(Lyapunov exponents and Kolmogorov entropy) seem to find a simpler
explanation than with the euclidean metric.

As an example, where euclidean metric is not very adequate, let us consider
the Baker's map \cite{schuster}. The interval $\left[ 0,1\right] $x$\left[
0,1\right] $ is mapped to $\left[ 0,1\right] $x$\left[ 0,1\right] $.
Therefore, the distance between two points can't be larger than the distance
between two opposite corners in $\left[ 0,1\right] $x$\left[ 0,1\right] $.
Nonetheless, the Baker's map has got a Lyapunov exponent bigger than one.
Then, the distance between neighboring points grows exponentially in a
finite region of the phase space. In the euclidean space we would have to
define the distance in this case as the euclidean length of the shortest
path lying entirely within the region that has suffered the deformation \cite
{barnsley}. As any nontrivial norm is equivalent to the euclidean or any of
the p-adics (Ostrwski's theorem \cite{ostro}), it would be convenient to
measure the distance between points in the Baker's map with a p-adic metric.

An ultrametric space is a space endowed with an ultrametric distance,
defined as a distance satisfying the inequality

\begin{equation}
d(A,C)\leq Max\{d(A,B)+d(B,C)\}
\end{equation}

( $A,B$ and $C$ are points of this ultrametric space), instead of the usual
triangular inequality, characteristic of euclidean geometry

\begin{equation}
d(A,C)\leq d(A,B)+d(B,C)
\end{equation}

\bigskip A metric space $\Bbb{E}$ is a space for which a distance function $%
d(x,y)$ is defined for any pair of elements $(x,y)$ belonging to $\Bbb{E}%
\frak{.}$

A norm satisfying

\begin{equation}
\left\| x+y\right\| \leq \bigskip \max \left\{ \left\| x\right\| ,\left\|
y\right\| \right\} ,  \label{3}
\end{equation}

is called a non-Archimedean metric, because equation \ref{3} implies that

\begin{equation}
\left\| x+x\right\| \leq \bigskip \left\| x\right\|  \label{4}
\end{equation}

holds, and equation \ref{4} does not satisfy the Archimedes principle:

\begin{equation}
\left\| x+x\right\| \geq \bigskip \left\| x\right\| .
\end{equation}

A metric is called non-Archimedean or ultrametric, if

\begin{equation}
d(x,z)\leq \max \left\{ d(x,y),d(y,z)\right\} .
\end{equation}
A non-Archimedean norm induces a non-Archimedean metric:

\begin{equation}
d(x,z)=\left\| x-z\right\| \leq \max \{d(x,y),d(y,z)\}.  \label{7}
\end{equation}

Equation \ref{7} implies a lot of surprising facts, e.g., that all triangles
are isosceles or equilateral and every point inside a ball is itself \ at
the center of the ball, furthermore the diameter of the ball is equal to its
radius.

An example of ultrametric distance is given by p-adic distance, defined as 
\begin{equation}
d_{p}(x,y)=\left\| x-y\right\| _{p}
\end{equation}

where the notation defines the p-adic absolute value: 
\begin{equation}
\left\| x\right\| _p=p^{-r},
\end{equation}

where $p$ is a fixed prime number , $x\neq 0$ is any integer, and $r$ is the
highest power of $p$ dividing $x$.Two numbers are p-adically closer as long
as $r$ is higher, such that $p^r$ divides $\left\| x-y\right\| $. Amazingly,
for $p=5,$ the result is that 135 is closer to 10 than 35.

Any positive or negative integer can be represented by a sum

\begin{equation}
x=\sum_{i=0}^\infty a_ip^i,
\end{equation}

where 
\begin{equation}
0\leq a_i\leq p-1.
\end{equation}

If negative exponents are considered in the sum, rational numbers can also
be represented. Such a representation is unique. The set of all sums $Q_p$
is the field of p-adic numbers, and contains the field of rational numbers $%
Q $ but is different from it.

\section{Lyapunov exponent and Kolmogorov entropy}

With the above description the p-adic numbers have a hierarchical structure,
whose natural representation is a tree. Let us now use this description to
work with the Bernoulli map (See \cite{schuster}).

In this way, the numbers can be represented as a set of points in a straight
line or by a hierarchical structure, depending on the definition of distance
(euclidean or Archimedean).

Let us represent  the initial value (state) to be mapped into the unit
interval by the sequence $0,a_1,......a_N$, with $a_i$ = 0 or 1 to denote
the initial value in binary notation. 

It is possible to reorder these sequences as a hierarchical tree. To get it,
let us do the following process to represent the result of the application
of the Bernoulli map:

We begin at an arbitrary point. We read, consecutively, the values of $a_i$
, from i=1 to N, of the sequence $0,a_1......a_N$. When $a_i$ takes the
value 0 we move to the left, and the same distance down. When $a_i$ takes
the value 1 we do the same, but moving on the right. The result is 2$^{N%
\text{ \ }}$branches of a hierarchical tree. Any branch represents
univocally a possible sequence $0,a_1......a_N$. Thus, for instance, the
sequence 0,0110 represent: left, right, right, left. Later, we will make $%
N\rightarrow \infty $ to recover the Bernoulli map.

The distance $d(x_{i},x_{j})$ between two branches (sequences) $x_{i},x_{j}$
in this tree is given by

\begin{equation}
d(x_i,x_j)=\left\{ 
\begin{tabular}{l}
$2^{m-N},i\neq j$ \\ 
$0,i=j$%
\end{tabular}
\right. ,  \label{2b}
\end{equation}
where m is the number of levels one must move up the tree to find a common
branch linking $x_i$ and $x_j$ , and N is the number of levels (the length
of the sequence). This is equivalent to

\begin{equation}
d(x_i,x_j)=\left\{ 
\begin{tabular}{l}
$2^{-h},i\neq j$ \\ 
$0,i=j$%
\end{tabular}
\right. ,
\end{equation}

where h is the position of the last block $a_h$ in which $\ a_i$ ($%
i=1,....,h $) are common to the two sequences $x_i,x_j$. It means that the
numbers $x_i$ and $x_j$ are close up to the $h^{th}$ binary place. This
distance is an ultrametric one.

Now we consider the map, given by $f(0,a_1.....a_N)=0,a_2.....a_Na^{^{\prime
}}$, such that $a_i$, i=1,...,N moves one position to the left, and a new
figure $a^{^{\prime }}$ is born, given the accuracy of the descripition. In
the limit when N$\rightarrow \infty $ the Bernoulli shift will be retrieved.

To calculate the Lyapunov exponent it is necessary to know how neighboring
points $x_0+\epsilon $ and $x_0$ evolve during the Bernoulli map. Let $%
\epsilon $ be equal to 2$^{-h}(1+2^{-\delta _1}+2^{-\delta _2}+.....)>2^{-N}$%
, then the first different position between $x_0=0,a_1..a_{h-1}a_h...a_N$
and $x_0+\epsilon $ is $a_h$. Then, it is necessary to move up the \ tree  N
- h +1 levels from the bottom line to find the common branch in the position 
$a_{h-1}$ (obviously, the last common figure between $x_0$ and $x_0+\epsilon 
$). Then

\begin{equation}
d(x_0+\epsilon ,x_0)=2^{-h+1}
\end{equation}

and

\begin{equation}
d(f^n(x_0+\epsilon ),f^n(x_0))=2^{-h+1+n}
\end{equation}

because the iteration $f^n$ moves away the common branch n positions from
the bottom level.

To calculate the Lyapunov exponent it is necessary to express the
exponential growth of the distance between two neighboring points:

\begin{equation}
\lim_{n\rightarrow \infty }\lim_{\epsilon \rightarrow 0}2^{\lambda
n}\epsilon =\lim_{n\rightarrow \infty }\lim_{\epsilon \rightarrow
0}d(f^n(x_0+\epsilon ),f^n(x_0))
\end{equation}

Since the base for measuring the p-adic distance in our space is the number
2, in the preceding equation we have expressed the exponential growth with $%
2^{\lambda n}$ instead of $e^{\lambda n}.$

Replacing $\epsilon $ and $d(f^n(x_0+\epsilon ),f^n(x_0))$ in the preceding
equation we obtain

\begin{equation}
\lim_{n\rightarrow \infty }\lim_{h\rightarrow \infty }2^{-h}(1+2^{-\delta
_1}+2^{-\delta _2}+.....)2^{\lambda n}=\lim_{n\rightarrow \infty
}\lim_{h\rightarrow \infty }2^{-h+1+n}  \label{17}
\end{equation}

from \ref{17} it can be easily observed that $\lambda =1.$

Since the Lyapunov exponent in Bernoulli map is $\ln 2$\cite{schuster}, we
recover this results with p-adic metric, since $2^1=e^{\ln 2}$. That means
that each unit time interval implies a new doubling of branches in each node
of the hierarchical tree. Then, once a unit time interval has elapsed, the
number of levels one must move up the tree to find a common branch increases
in one. This result will be crucial to understand how the information is
lost  in the course of time.

The former explanation has stressed the simplicity of the new expression $%
2^1,$ in comparison with $e^{\ln 2}$, and its interpretation.

In unidimensional maps, as the one considered here, the Kolmogorov entropy
coincides with the Lyapunov exponent \cite{schuster}. The Kolmogorov entropy
expression is:

\begin{equation}
k=-\lim_{n\rightarrow \infty }\lim_{\tau \rightarrow 0}\frac 1{n\tau
}\sum_{i_{1...}i_n}p_{i_{1...}i_n}\lg _2p_{i_1...i_n},
\end{equation}

where $p_{i_{1...}i_n}$ is the probability to reach the $i_n$ state of the
system in the phase space following a given path $i_1i_2...i_n$ . It can be
seen that in our case this probability only depends on the final state $i_n$
because for each state there is just one path, i.e., that given by the
sequence $i_1i_2...i_n$. Besides, the number of states in the $n^{th}$ level
is $2^n$ , and $\tau $ is the time elapsed to pass from one state to a
successive one. The probability to occupy one of the $2^n$ states is $%
p_n=p_{i_1i_2...i_n}=\frac 1{2^n}$ and it results

\begin{equation}
k=\lim_{n\rightarrow \infty }\lim_{\tau \rightarrow 0}\frac 2{\tau 2^n}.
\end{equation}

But the distance between two successive states of the $n^{th}$ level is $%
2^{1-n}$, because they are common until the ($n-1)^{th}$ level. Since the
speed $v$ to pass from one sequence to the next is constant in Bernoulli
map, i.e., $v=\frac{2^{1-n}}\tau =1$ the time $\tau $ elapsed between these
two successives states is $\tau =2^{1-n}$. As expected $k=1,$ coinciding
with the Lyapunov exponent.

The Kolmogorov entropy measures the loss of information in the process. From
our representation this loss of information can be easily seen, since the
process of separation of trajectories is such that for any step the increase
of the distance between two points duplicates the number of branches through
which this increment can be reached. We are loosing information because we
don't know exactly the way we are separating two states.

On the other hand, we can see that in the ultrametric space the natural time
of the system is also ultrametric. The time of transition between two
sequences $x_i,x_j$ satisfies the same expression \ref{2b} \ as the distance
between $x_i,x_j$.

Besides, subsequent behavior of two states that separate in a given point in
the ultrametric space depends from the point in which this separation
occurs, revealing an aging effect. This effect will be treated in future
works.

\section{Conclusions}

It was verified that the Bernoulli map leads to a hierarchical structure in
the p-adic metric. With the ultrametric distance the Lyapunov exponent and
the Kolmogorov entropy acquire a simpler expression and a direct geometric
interpretation is supplied by the hierarchical structure. The p-adic metric
seems to be the natural metric of this map. The hierarchical structure
generates p-adic properties for the temporal evolution.

\begin{acknowledgement}
\begin{acknowledgement}
This work has been partially supported by Alma Mater prize, Havana
University and Ministerio de Educacion y Cultura, Spain. We acknowledge
helpful suggestions and comments by Alvaro Perea and support from LCTDI,
UNED.
\end{acknowledgement}
\end{acknowledgement}


\begin{thebibliography}{99}
\bibitem{mezard}  M\'{e}zard, M. , G. Parisi, N. Sourlas, G. Toulouse, and
M. Virasoro, Phys. Rev.

Lett. 1984 \textbf{52}, 1156 . Also M\'{e}zard, M. , G. Parisi, N. Sourlas,
G. Toulouse, and M. Virasoro J. Phys. (Paris) 1984 \textbf{45, }843 .

\bibitem{rammal}  Rammal, R., G. Toulouse and M. A. Virasoro Rev. Mod. Phys
1986 \textbf{58}, 3, 765-788.

\bibitem{brekke}  Brekke, L., G. O. Freund, Physics Reports\textbf{\ }1993%
\textbf{\ 233}, 1, 1-66 .

\bibitem{stanley}  Zapperi, S., K. B. Lauritsen and H. E. Stanley Phys. Rev.
Lett 1995 \textbf{75}, 22, 4071-4074 .

\bibitem{pelayo}  Garcia-Pelayo, R., I. Salazar, W. C. Schieve J. Stat. Phys
1993 \textbf{72}, 1/2, 167-187.

\bibitem{prl}  Sotolongo-Costa, O., Y. Moreno-Vega, J.J. Lloveras and J. C.
Antoranz Phys. Rev. Lett 1996 \textbf{76}, 1, 42-45 .

\bibitem{schuster}  H. G. Schuster \textit{Deterministic Chaos }%
Physik-Verlag 1984.

\bibitem{barnsley}  M. Barnsley \textit{Fractals Everywhere }Academic Press
1988.

\bibitem{ostro}  A. Ostrowski, Acta Math 1918 \textbf{41} , 271 .
\end{thebibliography}
\end{document}